\documentclass[9pt,twocolumn,twoside]{opticajnl}
\journal{opticajournal} % use for journal or Optica Open submissions

\setboolean{shortarticle}{true}
\usepackage{lineno}
\usepackage{pifont}

\title{Application of Reconfigurable All-Optical Activation Unit based on Optical Injection into Bistable Fabry-P\'{e}rot Laser in Multilayer Perceptron Neural Networks}

\author[1]{Jasna V. Crnjanski}
\author[2]{Isidora Teofilovi\'{c}}
\author[1,]{Marko M. Krsti\'{c}}
\author[1,*]{Dejan M. Gvozdi\'{c}}

\affil[1]{University of Belgrade - School of Electrical Engineering, Bulevar kralja Aleksandra 73, 11120 Belgrade, Serbia}
\affil[2]{Department of Electrical and Photonics Engineering, Technical University of Denmark, Kgs. Lyngby, 2800, Denmark}

\affil[*]{gvozdic@etf.bg.ac.rs}

\begin{abstract}
In this paper we theoretically investigate application of a bistable Fabry-P\'{e}rot semiconductor laser under optical-injection as all-optical activation unit for multilayer perceptron optical neural networks. The proposed device is programmed to provide reconfigurable sigmoid-like activation functions with adjustable thresholds and saturation points and benchmarked on machine learning image recognition problems. Due to the reconfigurability of the activation unit, the accuracy can be increased by up to 2\% simply by adjusting the control parameter of the activation unit to suit the specific problem. For a simple two-layer perceptron neural network, we achieve inference accuracies of up to 95\% and 85\%, for the MNIST and Fashion-MNIST datasets, respectively.
\end{abstract}

\setboolean{displaycopyright}{false} % Do not include copyright or licensing information in submission.

\begin{document}

\maketitle

\section{Introduction}

Advantages offered by photonics in terms of speed, parallelism, and power consumption have propelled research in the field of photonics processors as a promising technology for hardware accelerators and artificial neural networks \cite{Shastri2020}. Recent research efforts have yielded efficient implementations of optical matrix multiplication by using plane light conversion, ring resonators based WDM technology or meshes of Mach-Zehnder interferometers \cite{Zhou2022}. However, the weakly-interacting nature of photons, which inherently favors the realization of linear transformations related to multiply and accumulate section, introduces a significant challenge in designing optical nonlinear units. 

To date, various approaches for all-optical activation units have been proposed, including those based on semiconductor optical amplifiers \cite{mourgias-alexandris2019}, induced transparency and reversed saturated absorption \cite{Miscuglio2018}, photonic crystal Fano lasers \cite{Ramussen2020}, coherent parametric processes in thin-film lithium niobate waveguides \cite{Li2023}, injection-locked distributed feedback lasers \cite{Liu2023}, etc. Proper activation function affects the overall test accuracy, so an additional challenge is to introduce flexibility, by generating different responses and thereby adapting to various machine learning tasks \cite{Campo2022}. Due to their stronger and more tunable nonlinearities, several electro-optic activation units offering reconfigurability to a variety of nonlinear functions have been proposed \cite{Fard2020, Wililiamson2020, Campo2022, Papas2023, Dehghanpour2023}. 

However, in the all-optical domain this task is more demanding. The programmability of the nonlinear transfer function shape and threshold can be accomplished by thermo-optic tuning on a silicon photonics platform, as in the case of the cavity loaded Mach-Zehnder interferometers \cite{Jha2020} or add-drop micro-ring resonators providing for lower threshold (< 0.1 mW) and smaller device-footprint \cite{Yu2022}. Programmability is also achieved by tuning the wavelength of an input optical signal, as demonstrated in an integrated solution based on hybrid Ge/Si micro-ring resonators that, due to strong thermo-optic effect in germanium, offers a low threshold (< 1 mW), although at the expense of increased fabrication complexity \cite{Wu2022}. Low-loss reconfiguration of the nonlinear function profile is reported by changing the power and duration of injected optical pulse for controlling the state of phase changed material loaded in silicon micro-ring resonator \cite{Fu2022}. All these solutions tackle some of the fundamental challenges in satisfying the requirements of photonics neurons, as low threshold, low power consumption and nonlinear transfer function reconfigurability to a variety of profiles typically used in deep neural networks. However, since they are all based on passive devices, an on-chip gain for amplifying the signal between layers can be necessary to ensure cascadability.

In \cite{OL} we showed that injection-locked Fabry-P\'{e}rot laser diode (FP-LD) in bistable regime can provide variety of nonlinear activation forms, from sigmoid to PReLU-like, which can be selected and adjusted simply by changing the laser bias current or the input optical signal frequency. This paper theoretically confirms that FP-LD under optical injection can provide a reconfigurable all-optical activation unit for multilayer perceptron neural networks, capable to produce an optical output suitable to drive subsequent neurons, hence providing cascadability necessary for deep optical neural networks (ONNs). The training of the ONN is done offline on a regular computer, by using a TensorFlow framework to find the optimal weight matrices for a given cost function. These final weights are subsequently employed for testing the ONN model within TensorFlow. Additionally, they are used for testing based on the numerical simulation of optical signal propagation through the activation unit.

In Section II, we briefly review the reconfigurable all-optical activation unit \cite{OL}. In Section III, we demonstrate improved performance of ONN using adaptive activation on the benchmark tasks of classifying images from the MNIST and Fashion-MNIST datasets. In the final section we give our conclusions.  

\section{All-optical activation unit}

\begin{figure}[t]
\centering
\includegraphics[width=\linewidth]{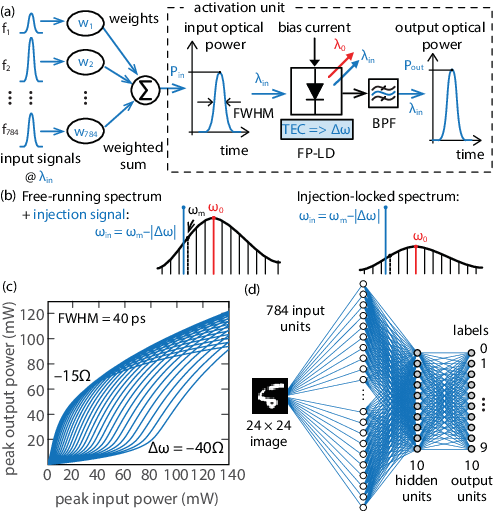}
\caption{a) Schematic of reconfigurable activation unit architecture based on FP-LD under optical-injection. b) Free-running spectrum of FP-LD under optical injection and in the case of injection-locking. c) Optical peak power transfer function of FP-LD for different values of $\Delta\omega$. d) Two-hidden-layer perceptron architecture.}
\label{fig:schematic}
\end{figure}

A schematic of proposed hardware realization of activation unit based on optical injection into FP-LD in bistable regime is presented in Fig. \ref{fig:schematic}(a) \cite{OL}. In the free-running (FR) regime, the dominant mode of the FP-LD operates at an angular frequency of $\omega_0$, corresponding to the central wavelength $\lambda_0$ (cf. Fig. \ref{fig:schematic}(b)). The input to the activation unit is an external optical pulse at the wavelength $\lambda_{\text{in}}$, (angular frequency $\omega_{\text{in}}$) which is injected in the FP-LD. The injection is intramodal, with a frequency detuning $\Delta\omega$ from the laser's nearest side mode, denoted as $m$, operating at an angular frequency $\omega_m$. To provide the bistable regime of FP-LD, it is necessary to keep frequency detuning negative ($\Delta\omega = \omega_{\text{in}} - \omega_m < 0$) \cite{JSTQE}. Due to injected optical pulse, the output spectrum of the FP-LD is altered to exhibit pronounced emission at the wavelength $\lambda_{\text{in}}$, which can be extracted by using a bandpass filter (BPF).

To model a FP-LD under optical injection, we utilize a system of multimode rate equations, accounting for optical injection with detuning $\Delta\omega$ in one of the side modes \cite{JSTQE}. The parameters of the rate equations correspond to the same structure, geometry, and material parameters of multi-quantum well active region as in \cite{JSTQE}. Under injection of a stream of short optical pulses into side mode $m$ with a negative frequency detuning $\Delta\omega$, the FP-LD can provide bistable regime and a nonlinear optical power transfer function, $P_{\text{out}} = \Phi(P_{\text{in}})$ at the wavelength of injected signal $\lambda_{\text{in}}$ \cite{OL}. The effect relies on dispersive bistability, which in the stationary case results in hysteresis loops in the space of input-output optical power corresponding to injected light $\lambda_{\text{in}}$, whose size and shape strongly depend on the FP-LD frequency detuning $\Delta\omega$ \cite{JSTQE}. In the dynamic (pulsed) regime, the hysteresis loop does not occur since injection-locking is not fully completed. In fact, the laser is in the transient regime, which is between injection-locked and unlocked state, and the optical transfer function exhibits abrupt variation with input power, which strongly depends on $\Delta\omega$, allowing for customization of the nonlinearity profile \cite{OL}. For example, in the case of small laser bias current of 1.1$I_{\text{th}}$, where $I_{\text{th}}$ = 8.2 mA represents the threshold current, and under intramodal injection of short Gaussian pulses with FWHM = 40 ps in the side mode $m$ = -9, a change in the frequency detuning magnitude $|\Delta\omega|$ from 15 to 40$\Omega$ ($\Omega = 10^{10}$ rad/s) produces a family of sigmoid-like profiles with threshold points, starting from below 1 mW and gradually increasing to over 70 mW (cf. Fig. \ref{fig:schematic}(c)). Additionally, the shape of the nonlinearity depends on the injected pulse width, pushing the threshold points toward larger optical power for decreased FWHM. Therefore, the device can serve as programmable all-optical activation unit that can provide reversible switching among different activation functions by using the input optical signal frequency detuning $\Delta\omega$ as a control parameter.  Since the signal power may drop below the activation threshold in the case of high optical loss of nonlinear layers with fixed response, the adjustable activation threshold together with the embedded gain represents a significant advantage.

\section{Machine learning benchmark}

For benchmarking the all-optical activation unit, we chose the MNIST and Fashion-MNIST image classification problems and use a two-hidden-layer perceptron architecture (cf. Fig.\ref{fig:schematic}(d)). The input 28$\times$28 greyscale images are normalized, reshaped into 784$\times$1 vector and  fed to the input layer. The 10-neuron output gives outputs corresponding to 10 different classes. The control parameter $\Delta\omega$ is independently selected for each of two layers ($\Delta\omega_1,\Delta\omega_2$), allowing for different activation functions across the layers.

The neural network model is implemented in the TensorFlow framework, with added Softmax layer output layer interpreting the results as a probability distribution. The model is trained to minimize the cross-entropy cost function using Adam optimizer, with learning rate 0.001 and batch size of 256 for 100 epochs. Overfitting problems are addressed by L2 weight regularization with parameter 0.02 in the second hidden layer, and a Dropout with 0.1 rate after the input layer. Since the activation function is an optical power transfer function, a non-negative constraint is applied to the neurons' weights and intercepts.

For training the ONN we use a custom-defined analytical function with coefficients $b_1,...,b_7$ as activation function:
\begin{equation}
P_{\text{out}}=b_1\ln{\{b_2+b_3\ln{[b_4+(\exp{(b_5P_{\text{in}}})-1)^{b_6}]}\}}+b_7P_{\text{in}}
\label{eq:fit}
\end{equation}
fitted to numerically calculated data points corresponding to complete family of transfer functions shown in Fig. \ref{fig:schematic}(c). 

\begin{figure}[t]
	\centering
	\includegraphics[width=\linewidth]{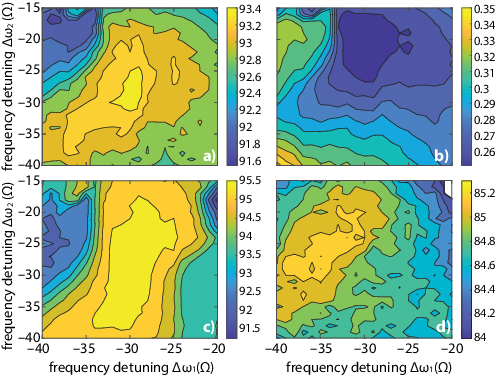}	
	\caption{Averaged testing a) accuracy and b) loss for MNIST and d) accuracy for Fashion-MNIST classification for all combinations of available activations in the hidden ($\Delta\omega_1$) and in the output layer ($\Delta\omega_2$). c) The same as a) for two-hidden-layer perceptron with 25 units in the hidden layer.}
	\label{fig:colormap}
\end{figure}

The MNIST dataset is divided into 60,000 training and 10,000 test samples (images) of handwritten digits (0-9). Model evaluation results for the test data are averaged over 10 independent simulation runs to compensate for differences in weight initialization. Fig.\ref{fig:colormap} shows colormaps for testing accuracy and loss for various combinations of available activations across the layers. The control parameter $\lvert\Delta\omega_1\rvert$ defines the activation function in the hidden layer and varies from 20 to 40$\Omega$. For the output layer, a slightly broader range for $\lvert\Delta\omega_2\rvert$ (from 15 to 40$\Omega$) can be used to achieve high accuracies. The network is found to converge to high accuracy (>91.5\%) for each combination of optical nonlinearities across the network, revealing results comparable with previous works \cite{Campo2022, Fard2020, Fu2022, Jha2020, Wu2022}. The highest achieved testing accuracy of 93.5\% is obtained for the detuning pairs (-29$\Omega$, -29$\Omega$) and (-29$\Omega$, -26$\Omega$), corresponding at the same time to the losses of 0.268 and 0.258, respectively. These values of detuning provide an activation profiles with thresholds between 15 and 20 mW (c.f. Fig. \ref{fig:schematic}(c)). Clear case in favor of using reconfigurable activation unit is reflected by the accuracy drop for other combinations of detuning. For example, for the detuning pair (-38$\Omega$, -15$\Omega$), the accuracy drops by almost 2\% to 91.54\% with a corresponding loss of 0.31. 

\begin{figure}[t]
	\centering
	\includegraphics{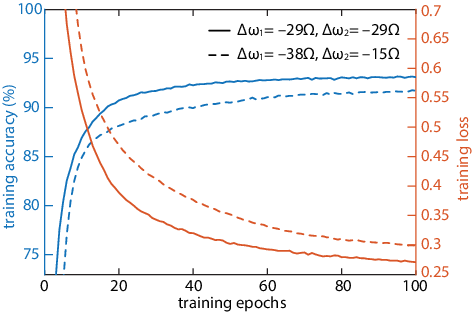}	
	\caption{The evolution of the accuracy (left axis, blue lines) and
losses (right axis, red lines) during the training of the model for activation pairs corresponding to high ((-29$\Omega$, -29$\Omega$), solid lines) and low (-38$\Omega$, -15$\Omega$), dashed lines) training accuracy.}
	\label{fig:training}
\end{figure}

Figure \ref{fig:training} shows the training accuracy and loss evolution for detuning combinations corresponding to the highest and lowest accuracy values for single training run without averaging. Presented optical activations reach training accuracies between 91-93\%. For training accuracies to exceed 90\% a minimum of 40 training epochs is needed, for any combination of activations. However, for optimal combination of activation functions, only 17 epochs are needed to reach training accuracy of 90\%.

\begin{figure}[b!]
	\centering
	\includegraphics[width=\linewidth]{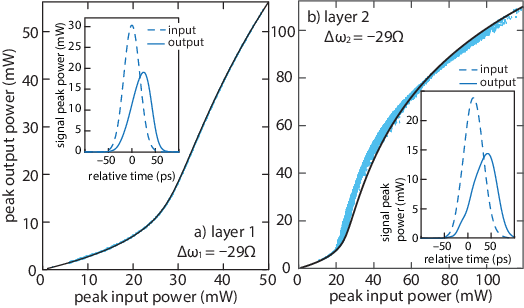}	
	\caption{Activations used for training (solid lines) and peak power values (blue dots) of actual optical signals propagating through (a) hidden and (b) output layer for ($-29\Omega,  -29\Omega$). Insets: examples of optical signals waveforms at the input (dashed lines) and output (solid lines) of activation units.}
	\label{fig:numerical}
\end{figure}

\begin{table*}
    \centering
    \small
     \caption{Accuracies per label for TensorFlow testing and for physical-layer simulation testing}
    %\resizebox{\columnwidth}{!}{
    \begin{tabular}{l c c c c c c c c c c}
    \midrule
Label &0	&1	&2	&3	&4	&5	&6	&7	&8	&9 \\
\toprule
TensorFlow&98.67	&97.71	&90.89	&91.78	&94.3	&89.01	&94.36	&93.29 &89.22	&88.90 \\
Physical simulation &98.37	&98.06	&89.44	&93.07	&93.08	&86.21	&95.82	&92.9	&88.6	&90.98
     \end{tabular}%}
   \label{tab:tabela}
\end{table*}

To further investigate applicability of the proposed activation function we perform testing of the pretrained ONN by implementing the activation unit physical model. In the numerical simulation, the input normalized data is represented as the peak power of optical Gaussian pulses with fixed FWHM = 40 ps. The signal multiplication by pretrained weights and summation is considered ideal, i.e., without signal degradation, while the response of each activation unit in the network is simulated by solving multimode rate equations \cite{JSTQE} for given input optical signal and frequency detuning $\Delta\omega$. Optical signals are multiplied with weight coefficients and after their summation, sent through the activation unit with control parameter $\Delta\omega_1=-29\Omega$ to obtain the output from the hidden layer. The peak power of output signals (blue dots) almost perfectly resemble the transfer function (black line) used for training (c.f. Fig. \ref{fig:numerical}(a)). However, these output signals do not preserve the ideal Gaussian shape of input signals and have slightly larger FWHM (on average around 45 ps) as shown in example given in the inset of Fig. \ref{fig:numerical}(a). Consequently, the numerical response of the output layer activation units deviates with respect to those used for training of the ONN, since activation units’ response used for training were calculated for ideal Gaussians with fixed FWHM. This leads to a dispersion of the peak power values (blue dots) around activation function used for training (black solid line) shown in Fig. \ref{fig:numerical}(b). It should be mentioned that transfer function used for training, shown in Fig. \ref{fig:numerical}(b) as solid line, is calculated for slightly broader input signals (45 ps instead 40 ps), since in this case discrepancy between numerically simulated optical pulses and transfer function is sufficiently small to preserve high network accuracy. The main reason for this adjustment of the transfer function used for training is the lack of simple and efficient physical simulation-based training procedure. Nevertheless, the classification accuracy calculated according to the peak power of optical signals at the output of neural network remains high and around 92.77\% which is just slightly lower than 92.9\% predicted by testing of neural network model in TensorFlow. High testing accuracy value obtained via physical-layer simulations, despite the obvious deviations of actual physical signals, indicates robustness of the activation unit against optical signal imperfections, which are typically introduced by fabrication processes or noise. A comparison of accuracies per label for TensorFlow and for physical-layer simulation testing is given in Table \ref{tab:tabela}.

To showcase the importance of activation function adaptivity to a task at hand, we perform the training of the TensorFlow ONN model with 25 units in the hidden layer. As expected, this leads to increased testing accuracy, as shown in Fig. \ref{fig:colormap}(c). In this case, the maximum accuracy of 95.84\% corresponds to frequency detuning pair (-31$\Omega$, -30$\Omega$). However, at the expense of using 15 neurons more than in the previous case, a large accuracy value (>95\%) can be obtained for wide range of activations in the second layer providing that $|\Delta\omega_1|$ is between 28$\Omega$ and 33$\Omega$. 

Finally, we tested performance of two-hidden-layer perceptron with 10 units in the hidden layer with Fashion-MNIST dataset that contains labelled apparel with more diverse and intricate patterns compared to MNIST images, leading to increased classification complexity. The classification accuracy for the same hyperparameters used for training as in the MNIST case reaches maximum value of 85.3\% after averaging 10 independent repeated runs (c.f. Fig. \ref{fig:colormap}(d)). However, high accuracy is achieved for hidden layer activations corresponding to larger values of detuning magnitude ($|\Delta\omega_1| > 32\Omega$) and thus larger activation threshold (c.f. Fig. \ref{fig:schematic}(c)). The range of activation functions that yield high accuracy values differs from those observed in the previously analyzed MNIST dataset. For instance, the activation pair (-38$\Omega$, -29$\Omega$) can achieve high accuracy with the Fashion-MNIST dataset, whereas with the MNIST dataset, the accuracy is approximately 3\% lower than the maximum.

\section{Conclusion}

In this paper, we propose the concept of all-optical multilayer perceptron neural network, in which activation units exploit optical injection into bistable Fabry-P\'{e}rot lasers. The unit can be dynamically reconfigured by programming the frequency detuning between input optical signal and side-mode under injection. Our simulations confirm that the ability to adjust the activation function profile provides a powerful tool for achieving high performance of NN, since the activation functions resulting in higher accuracies change depending on the addressed problem and optical networks architectures. In other words, adjusting the threshold and saturation points of activation function could allow for using the simpler and power efficient ONN architectures, while maintaining the performance quality. This innovative activation unit allows for the realization of strong nonlinearities without the requirement of having additional amplifiers between layers in a multilayer network.  

\begin{backmatter}
\bmsection{Funding} This research was supported by Science Fund of the Republic of Serbia, Grant no \#7750121, All-optical Reservoir Computer Architecture based on Laser Bistability – ORCA-LAB, and partially by Serbian Ministry of Education, Science and Technological Development.

\bmsection{Disclosures} The authors declare no conflicts of interest.

\bmsection{Data availability} Data underlying the results presented in this paper are not publicly available at this time but may be obtained from the authors upon reasonable request.

%\bmsection{Supplemental document} See Data File 1 and Data File 2 for supporting content. 

\end{backmatter}

% Bibliography

\end{document}